\nonstopmode
\pdfoutput=1 
\documentclass[fleqn,usenatbib]{mnras}
\usepackage{newtxtext,newtxmath}
\usepackage[T1]{fontenc}
\usepackage{ae,aecompl}

\DeclareRobustCommand{\VAN}[3]{#2}
\let\VANthebibliography\thebibliography
\def\thebibliography{\DeclareRobustCommand{\VAN}[3]{##3}\VANthebibliography}

\usepackage{graphicx}
\usepackage{amsmath, bm}
\usepackage{times}
\usepackage[mathscr]{euscript}
\usepackage{subfigure}
\usepackage{xcolor}
\usepackage[normalem]{ulem}

\def\gsim{\mathrel{\raise.3ex\hbox{$>$\kern-.75em\lower1ex\hbox{$\sim$}}}}
\def\lsim{\mathrel{\raise.3ex\hbox{$<$\kern-.75em\lower1ex\hbox{$\sim$}}}}
\def\leq{\mathrel{\raise.3ex\hbox{$<$\kern-.75em\lower1ex\hbox{$-$}}}}

\makeatletter
\newcommand\footnoteref[1]{\protected@xdef\@thefnmark{\ref{#1}}\@footnotemark}
\makeatother

\hypersetup{colorlinks, citecolor = [rgb]{0,0.5,0} }

\title[kSZ-Velocity Analysis]{Peculiar Velocity Estimation from Kinetic SZ Effect using Deep Neural Networks}

\author[Y.~Wang et al.]{Yuyu Wang$^{1,2}$
\thanks{E-mail:yuyuwang@sjtu.edu.cn},
Nesar Ramachandra$^{3, 4}$,
\thanks{E-mail:nramachandra@anl.gov},
Edgar M. Salazar-Canizales$^{5,6}$,
Hume A. Feldman$^{1}$,
\newauthor
Richard Watkins$^{7}$,
and Klaus Dolag$^{8, 9}$
\\
$^{1}$Department of Physics \& Astronomy, University of Kansas, Lawrence, KS 66045, USA.\\
$^{2}$Department of Astronomy, School of Physics and Astronomy, Shanghai Jiao Tong University, Shanghai, 200240, China.\\
$^{3}$High Energy Physics Division, Argonne National Laboratory, Lemont, IL 60439, USA.\\
$^{4}$Kavli Institute for Cosmological Physics, University of Chicago, 5640 South Ellis Avenue, Chicago, IL 60637, USA.\\
$^{5}$Universidad de Sonora, Hermosillo, Mexico.\\
$^{6}$Department of Physics, University of Arizona, Tucson, AZ 85721, USA.\\
$^{7}$Department of Physics, Willamette University, Salem, OR 97301, USA.\\
$^{8}$University Observatory Munich, Scheinerstr 1, D-81679 Munich, Germany.\\
$^{9}$Max-Planck-Institut f\"ur Astrophysik (MPA), Karl-Schwarzschild Strasse 1, 85748 Garching bei M\"unchen, Germany.}

\date{Accepted XXX. Received YYY; in original form ZZZ}
\pubyear{2020}

\begin{document}
\label{firstpage}
\pagerange{\pageref{firstpage}--\pageref{lastpage}}
\maketitle

\begin{abstract}
The Sunyaev-Zel'dolvich (SZ) effect is expected to be instrumental in measuring velocities of distant clusters in near future telescope surveys. We simplify the calculation of peculiar velocities of galaxy clusters using deep learning frameworks trained on numerical simulations to avoid the estimation of the optical depth. The image of distorted photon backgrounds are generated for idealized observations using one of the largest cosmological hydrodynamical simulations, the Magneticum simulations. The model is tested to be capable peculiar velocities from future kinetic SZ observations under different noise conditions. The deep learning algorithm displays robustness in estimating peculiar velocities from kinetic SZ effect by an improvement in accuracy of about 17\% compared to the analytical approach.

\end{abstract}


\section{Introduction}
The Sunyaev-Zel'dolvich (SZ) effect \citep{SunZel1970, SunZel1972, SunZel1980} describes the process of Cosmic Microwave Background (CMB) distortion caused by the inverse Compton scattering of CMB photons off by electrons in galaxy clusters. The SZ effect has two contributions: thermal (tSZ) and kinetic Sunyaev-Zel'dolvich (kSZ) effect. The tSZ effect is caused by the random motion of hot electrons in the intra-cluster medium, while the kSZ effect is caused by the bulk motion of galaxy clusters. Therefore, the kSZ effect can be used in estimating peculiar velocities of galaxy clusters \cite[e.g.][]{RepLah1991, BhaKos2008, KasAtrKoc2009, ZhaFelJus2008, AtrKasEbe2012,  PlanckAde2013, SayZemGle2015, PlanckAgh2017, SoeSarGia2017, Hurier2017, KirSav2018}. However, the weak signal of the kSZ effect makes its detection very difficult. \citet{HanAddAub2012} first detected the kSZ effect from CMB maps with the Atacama Cosmology Telescope (ACT) through pairwise momentum estimator. Using similar methods, several groups have detected the kSZ effect in both real and Fourier spaces \cite[e.g.][]{Planck2015xxxvii, SoeFleSto2016, SugOkuSpe2017, CalBeaYu2017, LiMaRem2017}. In addition, some studies detected the kSZ effect by cross-correlating kSZ temperature map with density or velocity field \cite[e.g.][]{SchFerVar2015, HilFerBat2016}.  \citet{Planck2018liii} detected the kSZ effect through measurements of the CMB temperature dispersion. Furthermore, \citet{MitBerNie2017} discussed the ability of measuring the kSZ effect for individual clusters in the upcoming multi-frequency surveys. With the improvements in the kSZ measurement, the estimate of peculiar velocities using kSZ effect for individual clusters may become possible.

The peculiar velocity field is a powerful tracer of density fluctuations, which is generally studied through ensemble statistics such as bulk flows, velocity correlation functions, and the pairwise velocity statistics \cite[e.g.][]{BorCosZeh2000, WatFelHud2009, KumWanFelWat2015, WanRooFel2018}. The pairwise velocity statistics is the mean value of the peculiar velocity difference of galaxy pairs at separation $\textbf{r}$ and is a widely used approach to study the large-scale velocity field \cite[e.g.][]{FerJusFel1999, JusFerFel2000, FelJusFer2003, ZhaFelJus2008, HanAddAub2012, PlanckXXXVII2015}. Traditionally, estimating peculiar velocities using kSZ effect requires information about optical depth, which describes the integration of electron densities. However, the measurement of optical depth has errors and biases that may affect the estimate of peculiar velocities. \citet{LinAguBak2015} has an average uncertainty of the cluster optical depth around 31\%
and \citet{MitBerNie2017} forecasts a average uncertainty about 24\% in observations. In addition, using emission-weighted, which is not observable, rather than density-weighted temperature in measurements may lead to a biased optical depth estimation \citep{DiaBorMos2005,DolagSunyaev2013}. In simulations, the optical depth varies between models with and without star-formation and feedback \citep{FleBleFin2016,FleNagMcD2017}. The weak kSZ signal and  optical depth errors make the kSZ peculiar velocity calculation imprecise and difficult. Machine learning algorithms may provide a simpler and more accurate method for estimating kSZ peculiar velocities.

Machine learning algorithms are designed without explicit programming of the physical phenomena, instead perform complex analyses in a data-driven manner. Some of the machine learning methods including Gaussian processes, decision trees, nearest neighbor algorithms and support vector machines have been used in astrophysical context (see \citet{Dalya2019} for a recent overview). A subset of Machine learning algorithms is a class of extremely flexible statistical models called the Deep Neural Networks. These typically have large number of trainable parameters that can be optimized from abundant amount of data, while the relevant features are extracted automatically. Utilization of such deep learning methods is rapidly increasing due to the availability of data, advancements of computational architectures (such as the Graphic Processors, Tensor Processors and dedicated accelerators), and the development of accessible software libraries (such as \verb|TensorFlow|, \verb|Keras|, \verb|Torch| and \verb|JAX|). Specifically, the Convolutional Neural Networks (CNNs), where features of images are extracted hierarchically in various layers of the deep network, are powerful tools in image-based regression, classification, compression, and generation tasks. 

However, the model interpretability and explainability of deep learning methods remain to be areas of active research. The over-parametrized architecture of deep CNNs results in a difficult uncertainty quantification, and features important assessment and understanding of failure modes. The dependence on hyper parameter searches, optimal architectures, network initialization, and optimization routines also contributes to the cryptic nature of the results achieved from deep learning algorithms. Thus,- deep learning algorithms are often characterized as 'black box' inference techniques. 

Despite these caveats,  deep learning neural networks trained on sufficient amount of data outperform the traditional classification and regression techniques (shown in various comparison studies, for instance \cite{metcalf2019} in strong lensing detection problem). Specifically in deep CNNs, the low-, mid-, and high-level image features computed in the initial, middle, and final convolutional layers, respectively, are used to correlate inputs and targets in a highly efficient manner. This makes CNNs practical tools in image processing tasks, including astronomical applications. 

Learning the intrinsic characteristics of the dataset may be accomplished unsupervised where the training is unaccompanied by correct responses, e.g., in Generative models \citep{Ravanbakhsh2016,morningstar2018,he2019learning}. Alternatively, a supervised routine involves learning correct mapping during training. Supervised techniques for object identification have been applied in a broad variety of astrophysical problems including strong lensing image classifications \citep{Petrillo2017} and parameter estimations \citep{levasseur2017,Hezaveh2017,morningstar2018}, which have demonstrated improvements to predictive precision and inference speed compared to traditional inference techniques. 

Machine learning applications in Cosmological analyses frequently deal with  simulated data instead of observational data. This is in part due to the possible lack of large quantity of observational data. On the other hand, the ability of calibrating the forward model parameters is not robust enough to generate unbiased training data. 

In this paper, we use simulation data to test the feasibility of extracting peculiar velocities from kSZ effect by deep learning architectures. In section~\ref{sec:kSZ}, we describe the relation between the SZ effect and the peculiar velocity. In section~\ref{sec:simulation}, we introduce the simulation we used for generating training and validation data. In section~\ref{sec:CNN}, we display the CNN structure of the deep learning model. In section~\ref{sec:training}, we show predictions of our model and compare it with the analytical method. In section~\ref{sec:PV}, we exam the model predictions through the pairwise velocity statistics. In section~\ref{sec:observation}, we test the feasibility of the model to observations under noise conditions. In section~\ref{sec:conclusion}, we conclude this paper.

\section{Sunyaev-Zel'dovich Effect}
\label{sec:kSZ}

The relation between radial motions of galaxy clusters and the observed radiation temperature was first introduced by \citet{SunZel1980} with the Equation~\ref{eq:ksz-vel}, where $v_e$ indicates the velocity of electron along the line of sight, $v_c$ is the line of sight peculiar velocity of cluster, $\tau = \int \sigma_T N_e dl$ is the Thomson Scattering optical depth, $\sigma_T$ is the Thomson Scattering cross-section, and $N_e$ is the electron density.

\begin{equation}
\frac{\Delta T_{kSZ}}{T_{CMB}} = - \frac{1}{c}\int \sigma_T N_e v_e dl \simeq - \frac{\tau}{c}v_c
\label{eq:ksz-vel}
\end{equation}

On the other hand, the tSZ effect \citep{SunZel1970} is usually expressed by the Compton y parameter:
\begin{equation}
\frac{\bigtriangleup T_{tSZ}}{T_{CMB}}=yf(x), \ \ \  y = \int\frac{k_BT_e}{m_e c^2}\sigma_T N_e dl,
\label{eq:tsz}
\end{equation}
where $f(x)=xcoth\left( x/2 \right) -4 $ and $x$ is the dimensionless frequency given by $x=h\nu/(k_BT_{CMB})$.

Since the kSZ signal is independent of the redshift and has a strong suppression on the secondary CMB anisotropy, the kSZ effect can be available up to the era of reionization. However, due to the weakness of the signal and the error in optical depth measurement, the peculiar velocity estimation from kSZ effect is very challenging in real observations. 

Alternatively, the potential of utilizing numerical simulations for estimating peculiar velocity from the kSZ effect are being studied extensively. For instance, \citet{SoeSarGia2017} has shown promising results with obtaining pairwise velocity statistics with kSZ effect by applying map filtering to the signals and used tSZ effect to estimate the average optical depth. 

For both observations and simulations, the requirement of optical depth estimation is inevitable when using the analytical method to calculate the kSZ peculiar velocity. In addition, the estimation of optical depth in simulations varies between models with and without star-formation and feedback. The measurement of optical depth for a single cluster in observation is even more challenging. Therefore, a method that can predict peculiar velocities from kSZ effect independently of the optical depth would reduce the difficulty in calculating kSZ peculiar velocities significantly. Deep learning algorithm provides a possible approach to achieve it. A training dataset from a numerical simulation with a realistic SZ map-making pipeline may empower the deep learning model to simplify the computation in the estimation of peculiar velocities by avoiding the map filtering and optical depth estimation. 

\section{Simulation and training data}
\label{sec:simulation}

Deep Neural Networks typically utilize a large amount of training data in order to capture the complexities in the data and optimize the model. Therefore, cosmological simulations that can provide a large number of galaxy cluster samples are necessary. In addition, the simulation data must resemble idealized observations from telescopes, which leads to a lightcone pipeline to generate kSZ and tSZ images.

In this paper, we use the Magneticum Simulations\footnote{\label{note1}http://www.magneticum.org} to generate kSZ and tSZ cluster images. The Magneticum simulations are a set of cosmological hydrodynamical simulations with a large range of scales and resolutions. The Magneticum Simulations are generated by an extended version of the N-body/SPH \verb|GADGET3| code \citep{SprYosWhi2001, Springel2005, BecMurArt2015} with WMAP7 \citep{WMAP7} cosmological parameters from \citet{KomSmiDun2011}. The dark matter only simulation includes dark matter and dark energy that provide gravity information, while the hydrodynamical simulation uses the hydrodynamic equations to include the baryonic component, which can be described as an ideal fluid. In addition, these simulations follow a wide range of physical processes \citep[for details, see][]{HirDolSar2014, TekRemDol2015}, which are important for galaxy formation and the evolution of the intra-galactic and intra-cluster medium \citep[see][and accompanying results]{BifDolBoh2013a, DolKomSun2016, GupSarMoh2017}. With the baryonic particles and temperature information, the SZ signal can be detected by tracking back along the line of sight. 

In this paper, we use the largest box, Box0 \citep[see also][]{BocSarDol2016, SoeSarGia2017, RagDolMos2019}, in the Magneticum Simulations. Table~\ref{T_Mill_ksz} shows the cosmological parameters of the simulation box and the parameters of our datasets. We take four redshift slices from the simulation that cover redshift in a range of [1.04, 2.15]. From those four redshift slices, we generated 40,000 kSZ and 40,000 tSZ images (10,000 images from each redshift slice) through \verb|SMAC| \citep{DolHanRon2005}, which is a map making utility for idealized observations. The size of a kSZ/tSZ cluster image is set to be twice its Virial radius, which is the radius within where the system obeys the Virial theorem. To reduce the calculation expense, we use the redshift of each slice instead of the redshift of each cluster in calculating the Virial radius, which means the size of the cluster images is not perfectly normalized to the Virial radius. According to our test, the difference is small and its effect on the final results is negligible.

\begin{table}
\caption{Specifications of the training data along with the cosmological parameters of the Magneticum Simulation Box0.}
\centering
{
\begin{tabular}{lc}
\hline
Matter density, $\Omega_m$ & 0.272\\
Cosmological constant density, $\Omega_\Lambda$ & 0.728\\
Baryon density, $\Omega_b$ & 0.046\\
Hubble parameter, $h$ ($100 km s^{-1} Mpc^{-1}$) & 0.704\\
Amplitude of matter density fluctuations, $\sigma_8$ & 0.809\\
Primordial scalar spectral index, $n_s$ & 0.963\\
Box size ($h^{-1}$Mpc) & 2688\\
Number of particles & $2\times4536^3$\\
Mass of dark matter particles, $m_{dm}$ ($10^{9} h^{-1} M_{\odot}$) & 13\\
Mass of gas particles, $m_{gas}$ ($10^{9} h^{-1} M_{\odot}$) & 2.6\\
Softening of particles, $f_p$ ($h^{-1}kpc$) & 10\\
Softening of stars, $f_s$ ($h^{-1}kpc$) & 5\\
\hline
Redshift range for clusters in slice 1 & [1.04,1.32]\\
Redshift range for clusters in slice 2 & [1.32, 1.59]\\
Redshift range for clusters in slice 3 & [1.59, 1.84]\\
Redshift range for clusters in slice 4 & [1.84, 2.15]\\
Mass of galaxy clusters & $[1, 70]\times10^{13} M_{\odot}$\\
Average mass of galaxy clusters & $10^{14} M_{\odot}$\\
Number of kSZ maps of each slice & $10,000$\\
Number of tSZ maps of each slice & $10,000$\\
Size of maps & $2R_{vir}$\\
\hline
\end{tabular}%
}
\label{T_Mill_ksz}
\end{table}

\begin{figure*} 
\includegraphics[width=17cm]{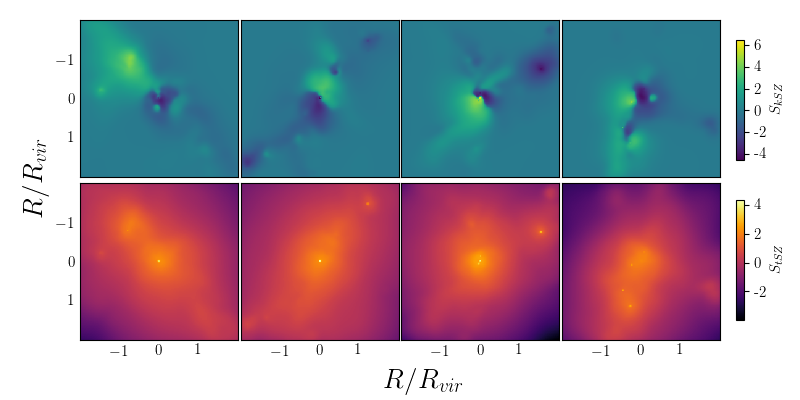}
\caption{kSZ (upper panels) and tSZ images (lower panels) of four clusters. $S_{kSZ}$ and $S_{tSZ}$ indicate the kSZ and tSZ signals re-scaled to increase the image contrast. For the kSZ signals in the upper panels, the color corresponding to the kSZ effect is presented via $S_{kSZ} = sinh^{-1}(\Delta T_{kSZ}/T_{CMB} \times 10^6)$. For the lower panels, the tSZ signal is a function of the Compton $y$ parameter, $S_{tSZ} = log_{10}(y \times 10^6)$. The width of each image equals to four times of the cluster Virial radius.}
\label{fig:pred}
\end{figure*}

Figure~\ref{fig:pred} shows the kSZ and tSZ examples of four clusters generated from the Magneticum Simulations. We train the neural network with 80\% of the images, which are similarly as the examples shown in figure, and use the rest 20\% of them as validation data for testing.

\section{The deep learning model}
\label{sec:CNN}

A custom-designed deep learning algorithm is implemented here to predict the peculiar velocity from kSZ effect. Convolutional Neural Networks (CNNs) are an obvious choice for such image-based regression analyses due to the following reasons: First, the amount of generated data (40,000 kSZ images) can be efficiently utilized in deep learning neural networks which consist of a large number of trainable model parameters called weights. It can be seen that with respect to the scaling of accuracy with the size of the dataset, deep learning neural networks outperform most existing machine learning models. Secondly, despite having characteristic features in the SZ signal (as seen in Figure \ref{fig:pred}), the feature-mapping to peculiar velocities is not straightforward due to the optical depth. This makes feature-agnostic training algorithms like CNNs more desirable than feature-specified learning methods for modeling SZ images. The CNNs can extract high and low level features from a series of convolutional filters, which are used to train the peculiar velocity prediction.

Numerous deep learning neural network architectures are currently in literature and under active research. However, we do not wish to compare different CNN variants in this work, nor claim to achieve the best possible accuracy in estimating peculiar velocities. our goal in this paper is  to demonstrate the feasibility of using deep learning neural networks to estimate peculiar velocities using the direct input of kSZ images and highlight the advantage of such simulation-based training approaches over the analytical calculation techniques on the kSZ peculiar velocity estimation.

\begin{figure} 
\centering\includegraphics[width=8.5cm]{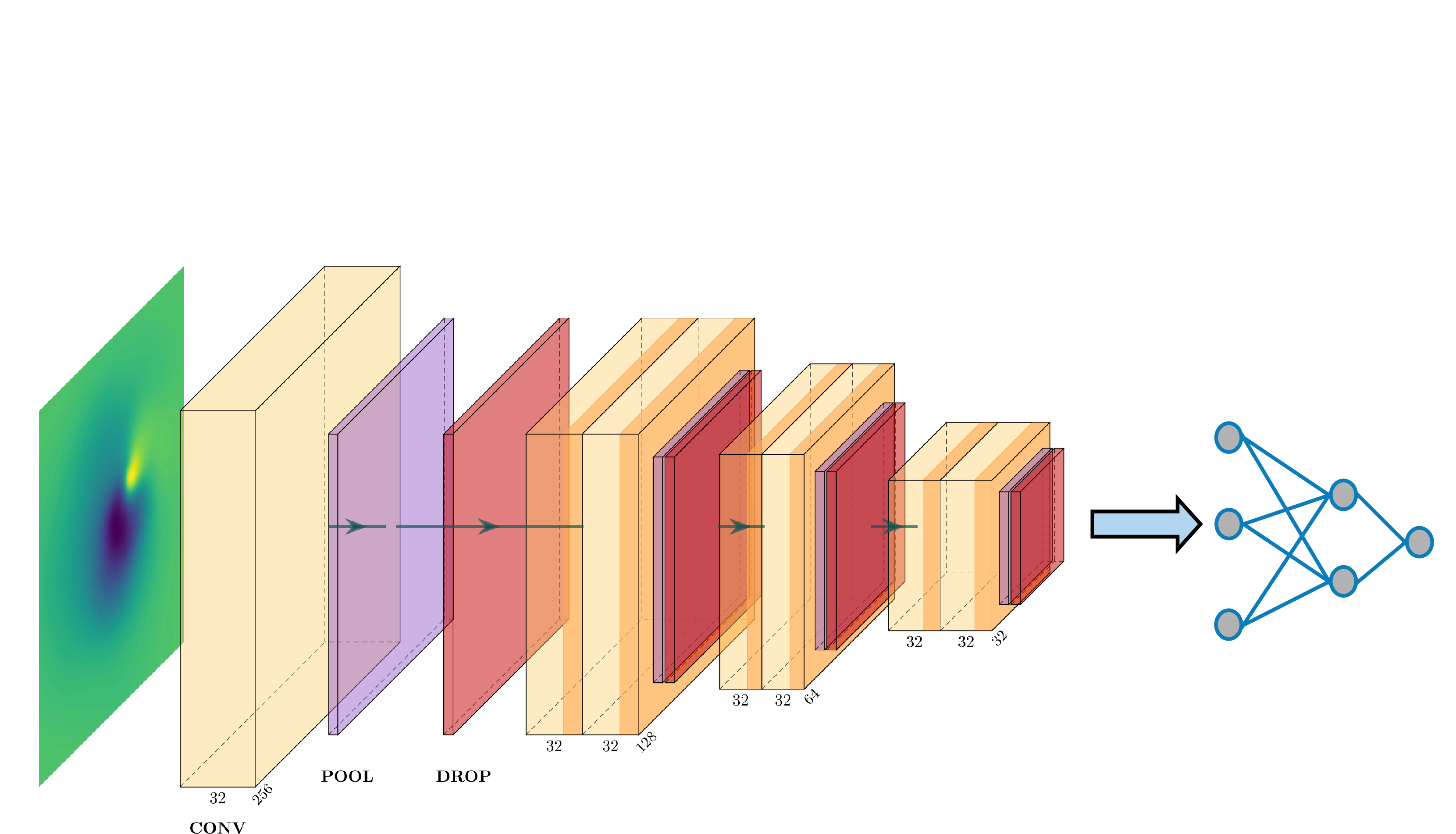}
\caption{Schematic Convolutional Neural Network architecture for regression including the kSZ effect only. The real architecture used in this analyses is multiple blocks of convolutional, pooling and dropout layers repeated before feeding the dense layers.}
\label{fig:cnn_ksz}
\end{figure}

\begin{figure} 
\centering\includegraphics[width=8.5cm]{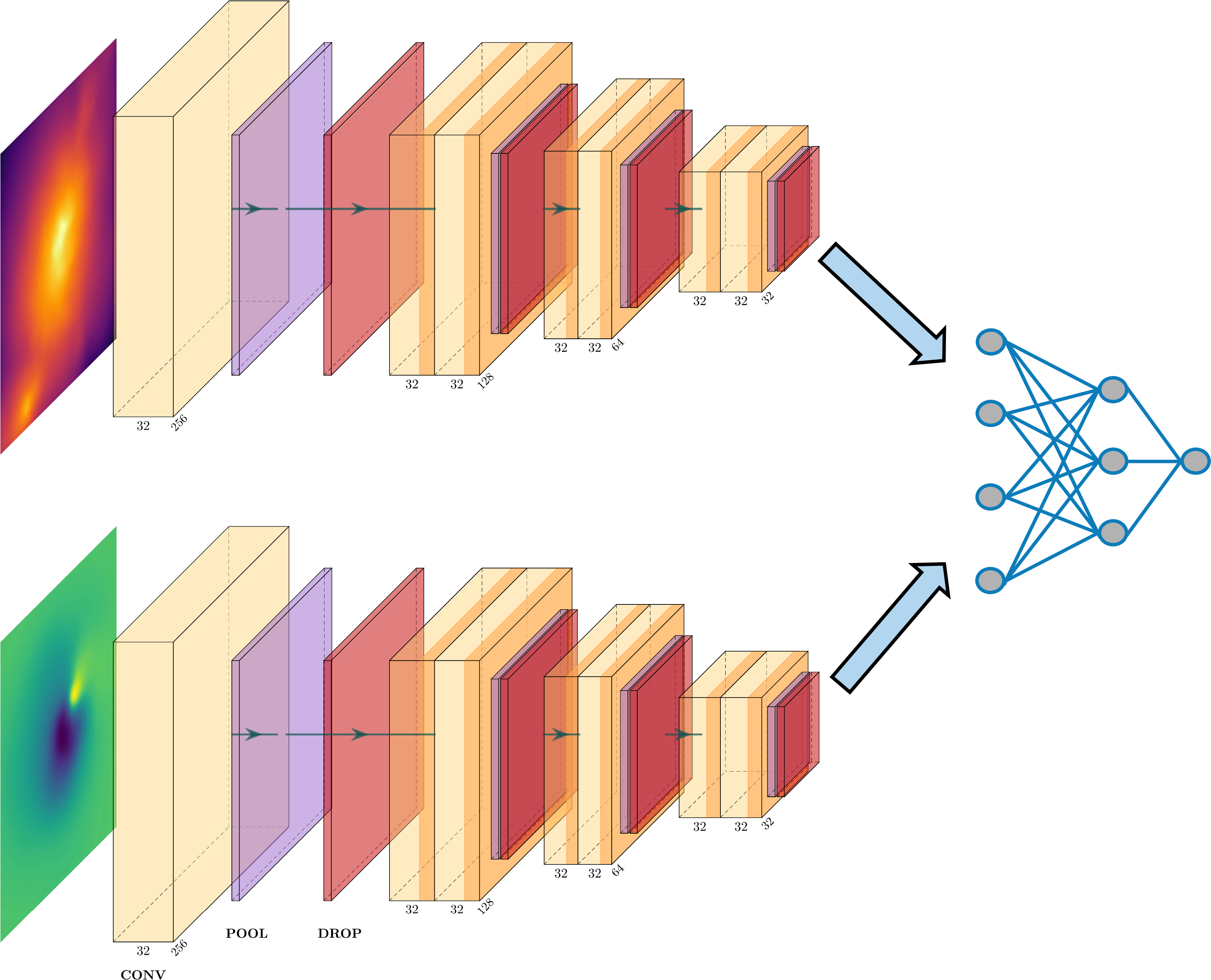}
\caption{Same as Fig.~\ref{fig:cnn_ksz} but for the both kSZ and tSZ effects. Two separate CNN branches process the images with input kSZ and tSZ signals, and the outputs from both the branches are then combined.}
\label{fig:cnn_ktsz}
\end{figure}

Figure \ref{fig:cnn_ksz} shows our CNN architecture with only the kSZ image as input data. It follows a Conventional Deep Neural Network architecture like the CIFAR-10 \citep{lecun2015deep}, with layers stacked sequentially. The kSZ image, the input data, will be addressed through several layer blocks (including convolutional, pooling and dropout layers) and multiple dense layers to get the peculiar velocity as the output. Short descriptions of each layers are as follows:  
1) Convolutional layers consist of numbers of image kernels that extract morphological features of the image. While the high-level features are extracted at the initial convolutional layers, more abstract features are obtained later. 2) The pooling layer operates on each map independently, and progressively reduces the spatial size of the map to reduce the amount of computations in the network. 3) Dropout layers  re-initialize a sub-set of neurons of the network at every epoch of the training, which reduces the chances of over-training. 4) The flatten later converts the 2-D matrix to a single 1-D vector. 5) Dense layers use fully connected neurons\footnote{For given inputs ${\bm x}$, the output ${\bm y}$ of each neuron is expressed in terms of its non-linear activation function $\phi$, weights ${\bm W}$, and biases $\bm b$ as ${\bm y} = \phi({\bm W} {\bm x} + {\bm b})$. The trainable parameters $({\bm W}, {\bm b})$ of the model are optimized during the learning phase.} to map this 1-D vector to the peculiar velocity corresponding to the input image. 

Overall, the repeated convolutional layer blocks extract abstract-featured maps from the images, which are then used as inputs in the dense layers toward to the end of the network. As opposed to the image classification, this regression pipeline has a linear activation to get point estimation of the peculiar velocity. In addition, the loss function is defined by the mean square error (MSE) value  $L = (v - v_p)^2$, where $v$ is the peculiar velocity from simulation, taken as true values (true velocity) for the training, and $v_p$ is the predicted peculiar velocity from the deep learning model. By providing enough correct data to learn from, the model would be trained to project the input kSZ image to the output peculiar velocity.

The model of including both kSZ and tSZ images has similar architecture with an independent repeating convolutional structure, shown in figure \ref{fig:cnn_ktsz}. The only difference in the combined kSZ and tSZ image analysis is that the kSZ and tSZ are computed in separate branches. After the flatten layer, the outputs from those two branches are concatenated to a 1-D vector, which is then fed into dense layers for predicting the peculiar velocity. 

\subsection{Uncertainty Quantification}

One of the shortcomings of a traditional regression analysis with CNNs is that it lacks proper treatment for the uncertainty quantification. This stems from the vast number of trainable parameters in the CNNs, such as the ones shown in Figures \ref{fig:cnn_ksz} and \ref{fig:cnn_ktsz}. A complete understanding of the posterior (in our case, the probability distribution of target peculiar velocities for given input SZ images) becomes intractable due to the large number of statistical model parameters. Bayesian Neural Network frameworks using Monte Carlo or Variational Inference techniques have been explored for solving such inference problems but several of these methods are challenging due to computational expenses, lack of convergence or clear diagnostics. 

Alternatively, Monte Carlo Dropout method \citep[see][for a detailed review]{gal2015dropout}  offers a middle ground for approximating the prediction uncertainty within reasonable computational overload. This is done by the utilization of existing trained deep learning models with dropout layers in prediction of the error bars around the mean estimates. 

A dropout layer, as explained previously, is generally used in CNNs to avoid over-fitting in the training phase. However, they can also be used in the testing phase as an approximate sampling scheme for model parameters. It was also shown by \cite{gal2015dropout} that the MC dropout is a Bayesian approximation of Neural Networks to Gaussian processes, where the error modeling is formally defined.

The implementation of Monte Carlo Dropout is as follows: We consider an ensemble of neural networks (with ensemble size $N_{tot}$) of the same architecture, but only different from each other by a fraction (prescribed by the dropout rate $d$) of trained neurons that are re-initialized to a random value (or `dropped-out'). Using the base architectures shown in Figures \ref{fig:cnn_ksz} and \ref{fig:cnn_ktsz} with dropout rate $d$, we obtain this ensemble of $N_{tot}$ networks. Each of these networks in the ensemble provide a different point-prediction of the peculiar velocity.

When a validation image $\mathbb{I}$ is forward propagated through each network in the ensemble, they provide individual predictions $v^i_p (\mathbb{I})$, where $i=0, 1, \ldots  N_{tot}$. These individual predictions $v^i_p (\mathbb{I})$ are different from each other due the fact that a different fraction their network parameters are dropped-out. The mean of all the individual predictions is calculated as $\displaystyle  \langle v_p \rangle = \frac{1}{N_{tot}} \sum_i^{N_{tot}} v^i_p(\mathbb{I})$ and the variance as $\displaystyle \sigma_v^2 = \frac{1}{N_{tot}} \sum_i^{N_{tot}} [v^i_p(\mathbb{I}) - \langle v_p \rangle]^2$ respectively. These aggregate mean and variance will be considered as the uncertainty quantified prediction from the ensemble. 

Hence, the MC dropout is a simple prediction uncertainty quantification tool without any additional expensive computation tasks while training, unlike the Bayesian Neural Networks that explicitly define distributions in predictions \citep{kendall2017uncertainties}. In addition to providing uncertainty estimations, such ensemble methods can also monitor failure modes, i.e., the choice of network architecture and training schemes can be compared in terms of robustness of the results. 

For our implementation of both kSZ and combined kSZ and tSZ we utilize an ensemble of $N_{tot} = 100$ networks for our predictions. We also use a large dropout rate $d=0.5$ to test models for both the consistency and the robustness of our final predictions. For the same $N_{tot}$, we have observed a small decrease in the prediction uncertainty with reducing the dropout rate, but the mean does not vary significantly.

\section{Training}
\label{sec:training}

We build two models respective to the two CNN architectures in section~\ref{sec:CNN}: Model I, kSZ only model shown in Figure \ref{fig:cnn_ksz}; Model II, the combined kSZ and tSZ model shown in Figure \ref{fig:cnn_ktsz}. For training strategy, we check the universality of the model in different redshifts by training the models with data of single redshift slices and with data of multiple redshift slices (all the four redshift slices).

For Model I, we first train the model with kSZ images of each redshift slice, which means we train the model four times independently and each time use the 80\% of the 10,000 kSZ images of a single redshift slice; secondly, we train the model with the data of multiple redshift slices, which means the training dataset is the 80\% of the entire 40,000 kSZ images.

\begin{figure} 
\centering\includegraphics[width=8cm]{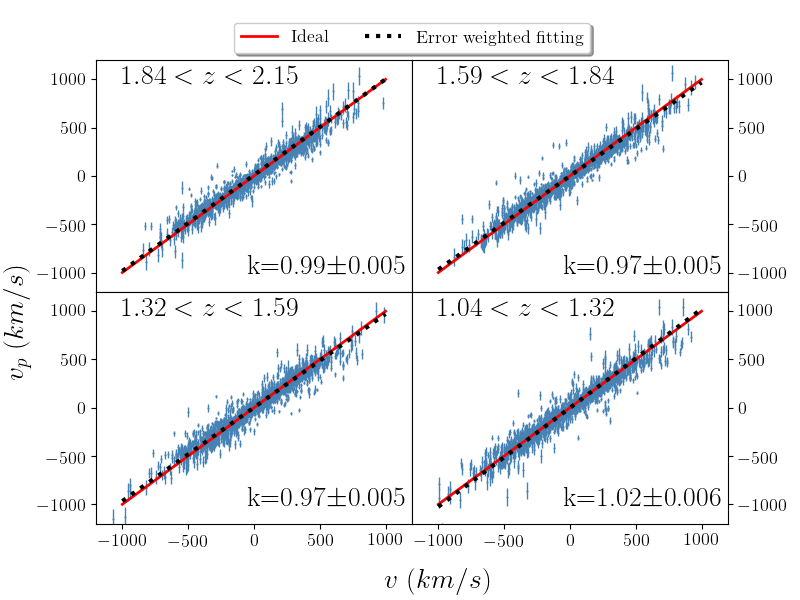}
\caption{The results of Model I trained by kSZ images in each redshift slice. The x- and y-axes show the true ($v$) and predicted  ($v_p$) peculiar velocities, respectively. The red solid line shows the 1:1 ideal relation between the true and predicted velocities, while the black dotted line shows the uncertainty weighted linear fitting of the scatter. The error bars show the uncertainty of the predicted velocity using the MC Dropout method.}
\label{fig:k_result_single}
\end{figure}

Figure~\ref{fig:k_result_single} shows the prediction results of Model I trained by kSZ images of single redshift slices. In the figure, the fitting line (black dotted line) is weighted by the uncertainty ($1/\sigma_v$), which represents the predictions accompanied by their error bars. The uncertainty weighted fitting result (black dotted line) agrees with the ideal expectation well, which is also represented by the fitting slope (k value). Although trained by data from different redshift slices, the prediction results of those four training sets have very similar fitting slopes, which means the model for predicting peculiar velocity from kSZ images is fairly stable with different redshifts. This is consistent with eq.~\ref{eq:ksz-vel} that the kSZ effect is independent of the redshift. In addition, the similarity of contours (tested but not shown in the figure) of the scatters of different redshift slices proves redshift independence.

\begin{figure} 
\centering\includegraphics[width=8cm]{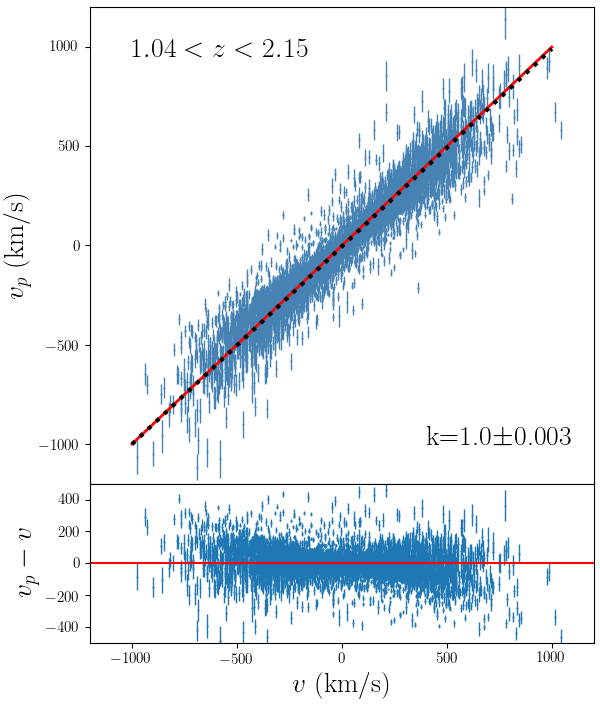}
\caption{The results of the Model I trained by kSZ images of multiple redshift slices. The x- and y-axes show the true ($v$) and predicted  ($v_p$) peculiar velocities, respectively. The red solid line shows the 1:1 ideal relation between the true and predicted velocities, and the black dotted line shows the uncertainty weighted linear fitting of the scatters. The error bars show the uncertainties of the predicted velocity using the MC Dropout method. The bottom panel shows the difference between predictions and expectations ($v_p - v$).}
\label{fig:k_result}
\end{figure}

Figure~\ref{fig:k_result} shows the results of the Model I trained by the kSZ images from multiple redshift slices, which covers a larger redshift region. Comparing with figure~\ref{fig:k_result_single}, figure~\ref{fig:k_result} has larger scatters due to more validation data. However, the fitting slope is similar to the one in figure~\ref{fig:k_result_single}. Though the model trained by the full data (from all four redshift slices) may have larger errors, it covers a larger region which makes the model more universal and flexible for applications.

In both figure~\ref{fig:k_result_single} and \ref{fig:k_result}, the predictions ($v_p$) using kSZ images show good agreements with the true velocities ($v$). However, the differences between the predictions and expectations cause scatter. Since the Model I does not include information of optical depth, we add tSZ information into the training to explore a possible improvement (Model II). However, the results of Model II show no significant difference from the results of Model I, which might lead to a conclusion that the deep learning neural network could estimate the peculiar velocity with only the kSZ input which simplifies the calculation significantly. The result of Model II is presented in Appendix~\ref{appendix:a}.

\subsection{Error analysis}
\label{sec:error}

In this section, we quantify the uncertainty in order to test the performance of our models. Since the difference between the results of Model I and Model II is negligible, we only present the error analysis for Model I in this section. In the figure~\ref{fig:k_result}, the MC dropout uncertainties (error bars of predictions) increases with the magnitude of the predicted velocity, and the average relative MC dropout uncertainty ($\sigma_v/v_p$) is about 25\%.

\begin{figure} 
\centering\includegraphics[width=8cm]{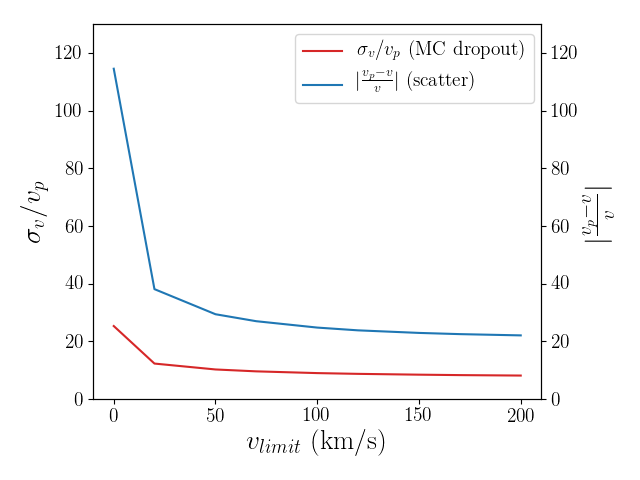}
\caption{The average MC dropout uncertainty and the average scatter in percentage of Model I trained by data from multiple redshift slices with different velocity limits. The x-axis is the velocity limits, for example, the label 20 in x-axis means eliminating all the predicted velocities lower than 20 km s$^{-1}$ ($|v_p|<20$). The y-axis shows the average MC dropout uncertainty $\sigma_v/v_p$ and the average relative scatter $\left|\frac{v_p - v}{v}\right|$ in percentage. The red line indicates the average uncertainty in percentage and the blue line shows the average scatter in percentage.}
\label{fig:k_E}
\end{figure}

However, the value of the relative MC dropout uncertainty is highly affected by its denominator, the predicted velocity $v_p$. Though the dropout uncertainty of low $v_p$ is smaller than the dropout uncertainty of high $v_p$, the smaller denominator will increase the relative uncertainty of low $v_p$. Therefore errors of low velocity clusters might bias the estimate of the average dropout uncertainty. We set different velocity limits ($v_{limit}$) to eliminate the effect from the low velocity, which is shown by the red line in figure~\ref{fig:k_E}. Eliminating velocities lower than 20 km s$^{-1}$ reduces the average uncertainty significantly to about 12\%. With larger velocity limits, the average uncertainty converges to about 8\%.

In addition, the scatter, which is the difference between the true velocity $v$ and the predicted velocity $v_p$, is another factor that affects the accuracy of the prediction. While the dropout uncertainty is a measure of precision (statistical uncertainty), scatter is a measure of accuracy (systematic uncertainty). In the bottom panel of figure~\ref{fig:k_result} (the residual plot), the absolute differences between predictions and expectations are mostly smaller than 200 km s$^{-1}$, which we define as the scatter. We also see an increasing trend of scatter with the magnitude of the velocity, but it is not a very strong dependence. Using the same method, we calculate the average relative scatter (difference of predictions and expectations over expectations) with different velocity limits, which is shown by the blue line in figure~\ref{fig:k_E}. After eliminating the velocities lower than 20 km s$^{-1}$, the average scatter becomes to about 38\%. With larger velocity limits, the average scatter converges to about 20\%.

\subsection{Comparison with analytical calculations of peculiar velocity}
\label{sec:Traditional}

The analytic calculation for estimating peculiar velocities from kSZ effect requires information of optical depth of each individual clusters (equation~\ref{eq:ksz-vel}). The optical depth of individual clusters in this paper is calculated through equation:
\begin{equation}
\tau_{cluster} = \frac{\int_0^{R_{vir}}\int_{l_-}^{l_+}\sigma_{T}N_edldr}{\pi R_{vir}^2},
\end{equation}
where $l_- = -100 h^{-1}$Mpc, $l_+ = +100 h^{-1}$Mpc and $R_{vir}$ is the Virial radius of clusters. Therefore, the optical depth of individual clusters is calculated by the averaging electron density within the Virial radius. The integral distance $dl$ for the optical depth is 200 $h^{-1}$Mpc, which is large enough to get a converged optical depth value. The kSZ value used in the analytical method is calculated by averaging the kSZ signals of each cluster within its Virial radius.

\begin{figure*} 
\centering\includegraphics[width=17cm]{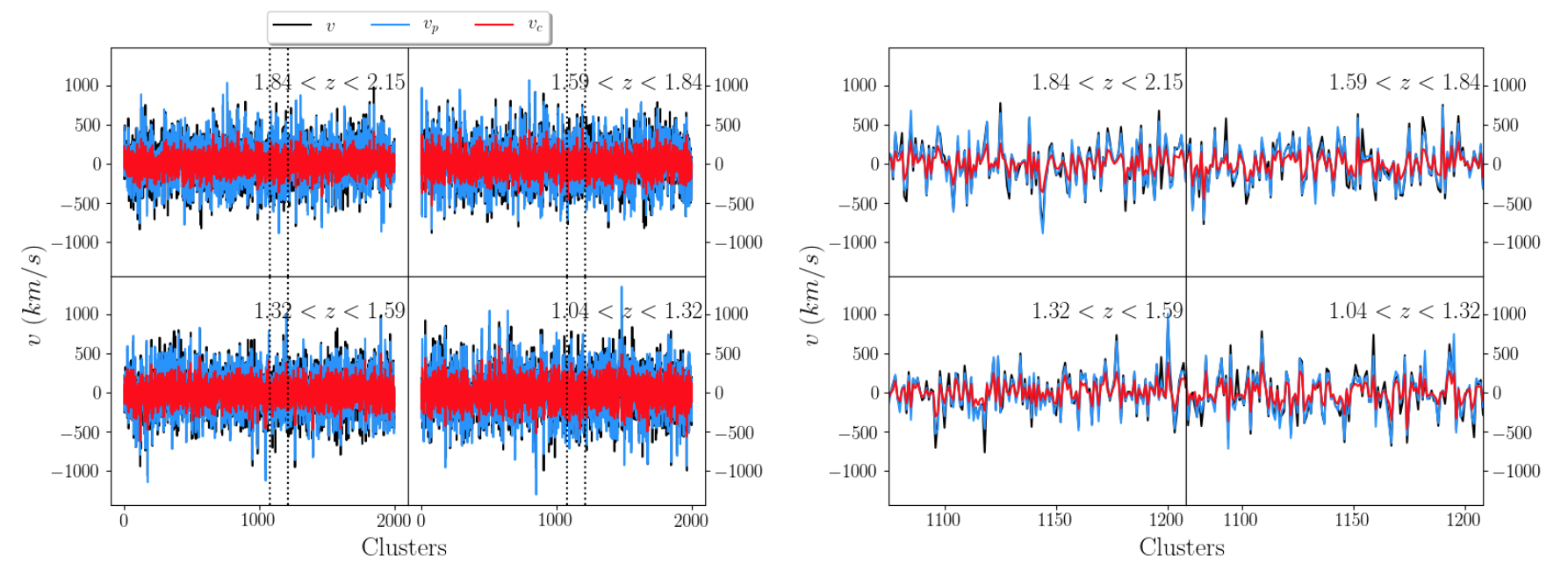}
\caption{The results of Model I (blue), the analytical method (red), and true velocities from the simulation (black). The four left panels show the result of 2,000 testing clusters in each redshift slice. The right panels show the amplification of the selected areas.}
\label{fig:v_compare}
\end{figure*}

\begin{figure} 
\centering\includegraphics[width=8cm]{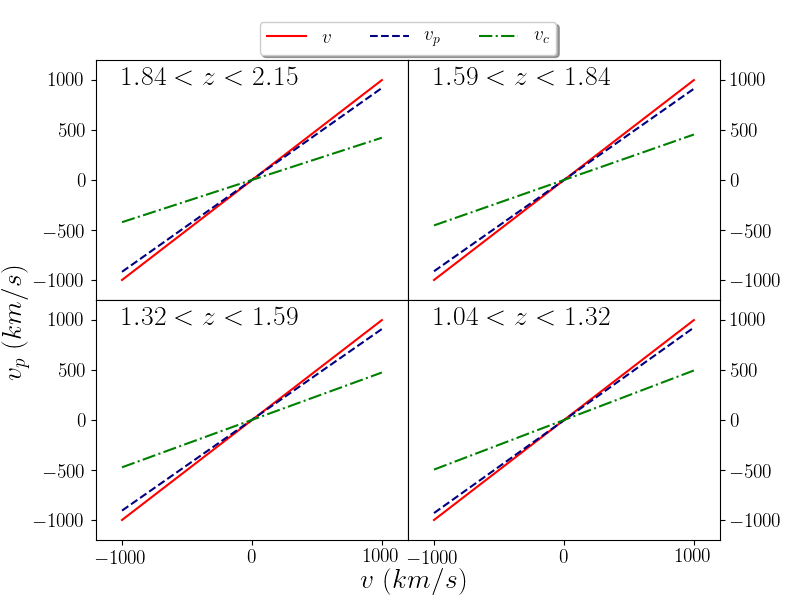}
\caption{Linear fittings of results of Model I and the analytical method. The x- and y-axes show the true ($v$) and predicted  ($v_p$) peculiar velocities, respectively. The red solid line shows the 1:1 ideal relation between the true and predicted velocities, the navy dashed line shows the linear fitting of Model I predictions, and the green dash-dotted line shows the fitting of the analytical result.}
\label{fig:k_v_compare}
\end{figure}

Figure~\ref{fig:v_compare} shows the results of Model I and the analytical method ($v_c$) of each redshift slice. From the figure, we could find that both the predictions using the deep learning neural network and the analytical method show strong correlations with the true peculiar velocity from the simulation. However, predictions of the analytical method have smaller magnitude than predictions of the deep learning neural network. The bias caused by the smaller magnitude becomes more obvious in the figure~\ref{fig:k_v_compare}. The fitting slope of the analytical method is around 0.45, which indicates a significant bias from expectation. Since predictions of the analytical method have no uncertainty, the fitting lines in figure~\ref{fig:k_v_compare} are not weighted by uncertainties. In section~\ref{sec:error}, the average scatter ($\left|\frac{v_p - v}{v}\right|$) of Model I is about 38\%, which is calculated with velocity limits of 20 km s$^{-1}$. Using the same method, the average scatter of the analytical method with 20 km s$^{-1}$ velocity limits is about 55\%. Comparing the average scatter of those two methods, the deep learning algorithm improves the accuracy of the velocity prediction by about 17\%.

For the analytical method, the choice of the averaging area of the cluster is heuristic, therefore, the calculations of kSZ signals and optical depth are affected by the averaging radius or aperture. We set the averaging radius of the calculation to be the Virial radius of each cluster. However, this choice of averaging radius may miss some features of kSZ signals outside that radius. The deep learning algorithm, instead, provides a better approach for dealing with the image that it can extract more details about the cluster pattern from kSZ images with convolutional neural networks. Therefore, it provides less biased velocity predictions. In addition, the deep learning neural network provides reasonable predictions without optical depth information. This makes it a more powerful tool for estimating peculiar velocities in observations due to the difficulty in measuring optical depth.  

\section{Pairwise Velocity}
\label{sec:PV}

Though our model trained by simulation data provides predictions with average uncertainties around 12\%, the average scatter off from expectations is not ideal (38\%). In addition, the uncertainty using observational kSZ signals can be worse due to difficulties in the detection. Therefore, ensemble statistics of peculiar velocities, rather than analysis of individual velocities, may be required. In this section we apply pairwise velocity statistics to our predicted peculiar velocities. 

\begin{figure} 
\centering\includegraphics[width=8cm]{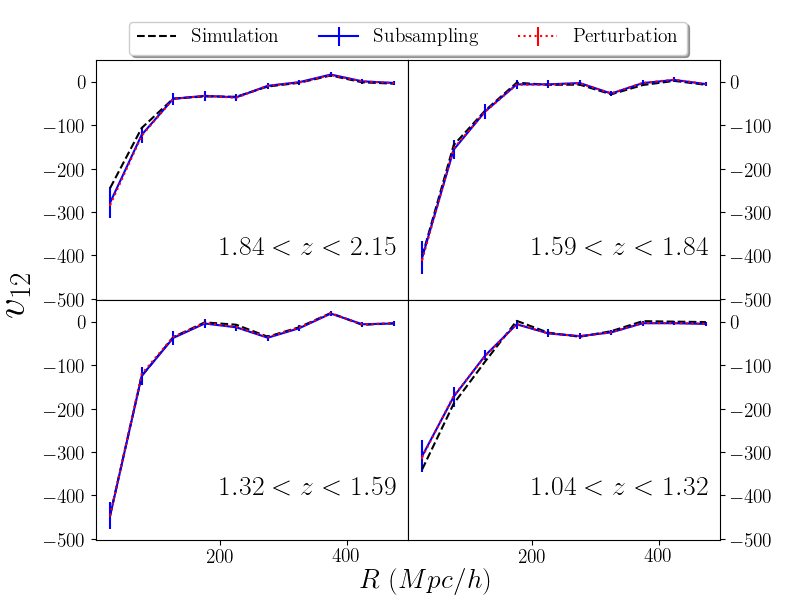}
\caption{The pairwise velocity estimates of the true and predicted velocities from Model I without velocity limits. The black dashed line shows the result of the true velocities, the blue solid line indicates the result of the predicted velocities with error bars calculated by subsampling method, and the red dotted line indicates the result of predicted velocities with error bars calculated by perturbation method.}
\label{fig:pw_v}
\end{figure}

Figure~\ref{fig:pw_v} shows the pairwise velocity statistics of Model I trained by the data from multiple redshift slices. In the pairwise velocity calculation, we use all of the predicted velocities without any velocity limits. Although the uncertainty and scatter without velocity limits are larger, the pairwise velocity of predictions agree with the result of the true velocities with small uncertainties (error bars). The uncertainties of the pairwise velocities are calculated in two different ways: 1)  subsampling method and 2) perturbation method that is perturbing the velocity catalog 100 times by the MC dropout uncertainty and calculating the statistical error through the standard deviation of the 100 perturbed catalogs. In the figure, the error bars of the perturbation method are so small, that they are invisible. 

\section{Adaptation to Observations}
\label{sec:observation}

To test the feasibility of our model to observations, we mimic observational kSZ signals by perturbing the simulated kSZ images with noise. We employ three types of noise in the perturbations: 1) Gaussian blur noise, 2) white noise, and 3) residual tSZ signals.

\begin{figure} 
\includegraphics[width=8cm]{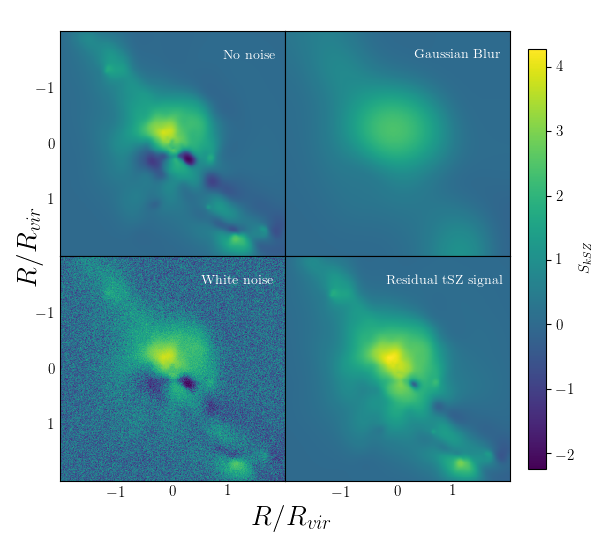}
\caption{kSZ images with different noise schemes. $S_{kSZ}$ indicates the re-scaled kSZ signal to increase the image contrast. The unit of $S_{kSZ}$ is $sinh^{-1}(\Delta T_{kSZ}/T_{CMB} \times 10^6)$.}
\label{fig:ksz_noise}
\end{figure}

Figure~\ref{fig:ksz_noise} shows the example images using different noise schemes. For 1) the Gaussian blur noise, we set the smoothing width as 40\% of the cluster Virial radius, thus its width varies between observations. Here, we do not use any observation setting as a reference for the Gaussian blur width, since we are only testing the effect of its noise on the model with simulation data. For 2) the white noise scheme, we use Gaussian noise with  standard deviation equals to the average value of the original kSZ signal, which means the signal-to-noise ratio equals  one. Again, this ratio is only used for testing. For 3) the residual tSZ signal, which is a source of error in kSZ detection, we added 10\% tSZ signals (from the simulation) to the kSZ image to mimic the possible noise caused by the remnant tSZ signals in kSZ observations. 

\begin{figure} 
\includegraphics[width=8cm]{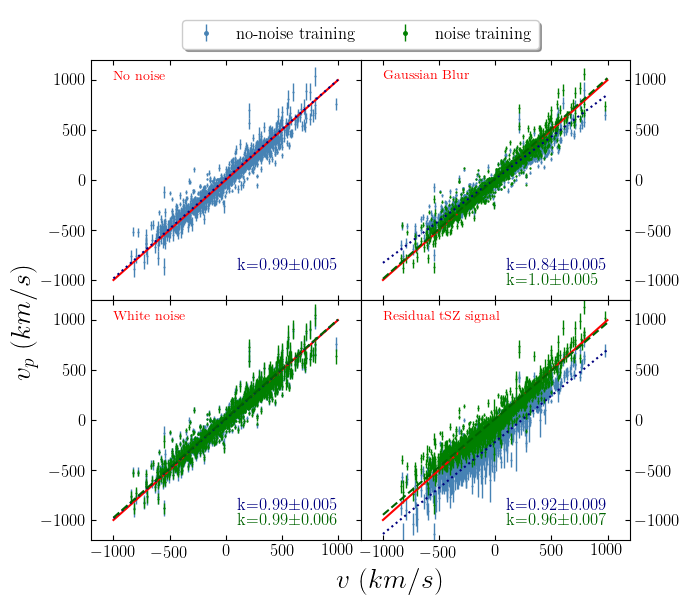}
\caption{Predictions with different noise schemes. The x- and y-axes show the true ($v$) and predicted ($v_p$) peculiar velocities, respectively. The models are trained by data of multiple redshift slices with (green) and without (blue) noise and tested by the noisy kSZ images. The red line shows the 1:1 ideal relation between the true velocity and the predicted velocity. The blue scatters and navy dotted lines show the results and uncertainty weighted fitting of models trained by kSZ images without noise and tested by the kSZ images with noise. The green scatters and dark green dashed line show the results and uncertainty weighted fitting of models trained and tested by noisy kSZ images.}
\label{fig:ksz_noise_test}
\end{figure}

We test our model with those noise schemes and present the results in figure~\ref{fig:ksz_noise_test}. One should notice that we implement two methods in the test: 1) the model is trained without noise but tested with the noisy images (blue scatters and navy dotted lines); 2) the model is both trained and tested  by the noisy images (green scatters and dark green dashed lines).

For method 1), our model shows great compatibility with the white noise. However, the prediction of adding residual tSZ noise shows biased results. We found that the bias caused by tSZ can be regarded as a constant shift from the ideal expectation. The larger the residual tSZ signal is, the larger the shift is. Therefore, the problem caused by the residual tSZ signal might be solved by given a correction. In contrast, the bias caused by the Gaussian blur drives the prediction magnitude smaller.

For method 2), the prediction of white noise shows no significant difference from method 1), while, biases caused by the Gaussian blur and residual tSZ noise are improved significantly by training the model with noisy images. The improvement shows the capability of our model on dealing with noisy kSZ signals.

Due to the difficulty in kSZ and optical depth measurements, the observational kSZ detection for individual clusters is very rare and the peculiar velocity estimated from kSZ observations is not accurate enough \citep{SayMonMro2019} to train a deep learning model. Therefore, the possibility of applying  deep learning neural network model to estimating peculiar velocities from kSZ observations is to train the model with perturbed simulation data. In this paper, we only tested three possible sources of uncertainties (Gaussian blur, white noise, and non-cleaned tSZ signal) in observations, and the noise intensities are set only for testing. A real kSZ observation may include different kinds of noise, such as noise from CMB anisotropies \citep{AghGorPug2001} and dusty star formation galaxies. According to \citet{MitBerNie2017}, the noise from CMB anisotorpies and residual tSZ would not be the dominant sources of uncertainty for the Cerro Chajnantor Atacama Telescope(CCAT). Instead, the kSZ detection will be significantly affected by the image resolution and emission from dusty star formation galaxies. Therefore, to apply deep learning algorithm to a specific observation (such as CCAT-prime), the simulated training dataset would have to include noise that represents the corresponding observational conditions, which will be studied in future work. Considering the advantages of using deep learning neural networks than using the analytical method (Sec.~\ref{sec:Traditional}), estimating peculiar velocities from kSZ effect with deep learning algorithms is very promising. With the upcoming kSZ detection telescopes such as CCAT-prime, a suitable machine learning model for observational kSZ is foreseeable.

\section{Conclusion}
\label{sec:conclusion}

The analytical method of estimating peculiar velocities from kSZ effect requires several steps, such as map filtering and optical depth calculation. In addition, the error in optical depth estimate makes it difficult to predict the peculiar velocity accurately.

In this paper, we test the feasibility of using deep learning neural networks to simplify the estimation of peculiar velocities from the kSZ effect. By comparing results of different redshift slices, using the simulation data, we find that our deep learning model is redshift independent, which is consistent with theory. 

Considering the relation between the tSZ effect and the optical depth, we build models that are trained by kSZ images (Model I) and kSZ+tSZ images (Model II). Those two models have similar predictions and uncertainties. We find that the average uncertainty of Model I is about 12\% and the average scatter is about 38\%. Although the average scatter is not ideal, the pairwise velocity of the predictions indicates that our model can provide reliable kSZ peculiar velocities for cosmological studies.

The similar results of Model I and Model II indicate that including tSZ does not show significant improvement to the velocity prediction when using deep learning neural networks. According to our model structures and prediction results, we have two explanations for the 'invalid' tSZ. First, using kSZ as the only input of the model to predict the peculiar velocity is enough. The predictions of Model I have a good fitting with slope close one. The improvement of adding tSZ effect is vanishingly small and redundant. In this circumstance,  simplifying the kSZ peculiar velocity estimate using deep learning neural networks performs even better than our original expectations. Second, our model structure and training is not accurate enough to show the effect of tSZ. Though the prediction of Model I has good fitting curve, adding tSZ input may have the potential to improve the scatter, which needs more detailed studies in the future. 

We tested the feasibility of our model to observations by perturbing the kSZ signals with three different noise schemes: Gaussian blur, white noise, and residual tSZ noise. For using simulation training data and noisy validation data, the prediction with white noise show few biases, while the biases caused by the Gaussian blur and residual tSZ noise are more significant. However, these biases can be improved by using noisy data for both training and testing. It clearly shows that deep learning neural network can be used to estimate peculiar velocities from the kSZ effect with both simulations and observations. A possible way for applying deep learning neural network to observations is to train the model with simulated training datasets that include noise corresponding to observations. However, a suitable model for observations still needs more kSZ detection of individual galaxy clusters in the future.

In conclusion, using deep learning neural networks to estimate peculiar velocities from the kSZ effect is both feasible and promising. It could simplify the analytical calculation of kSZ peculiar velocities significantly with only kSZ input, which avoids the estimation of optical depth as well as the map filtering.

\section*{Acknowledgements}

This work used the Extreme Science and Engineering Discovery Environment (XSEDE), which is supported by National Science Foundation grant number ACI-1548562.

NR's work at Argonne National Laboratory was supported under the U.S. Department of Energy contract DE-AC02-06CH11357. 

HAF and RW were partially supported by NSF grant AST-1907404. An award of computer time was provided by the INCITE program. This research used resources of the Argonne Leadership Computing Facility, which is a DOE Office of Science User Facility supported under Contract DE-AC02-06CH11357.  

KD acknowledge the support by the Deutsche Forschungsgemeinschaft (DFG, German Research Foundation) under Germany's Excellence Strategy - EXC-2094 - 390783311. 

The calculations were carried out at the Leibniz Supercomputer Center (LRZ) under the project 'pr86re'. We are especially grateful for the support by M. Petkova through the Computational Center for Particle and Astrophysics (C2PAP) and the support by N. Hammer at LRZ when carrying out the Box0 simulation within the Extreme Scale-Out Phase on the new SuperMUC Haswell extension system. NR thanks Jon\'as Chaves-Montero for discussions and help with the manuscript.

\bibliographystyle{mnras}
\bibliography{Yuyu}

\begin{thebibliography}{}
\makeatletter
\relax
\def\mn@urlcharsother{\let\do\@makeother \do\$\do\&\do\#\do\^\do\_\do\%\do\~}
\def\mn@doi{\begingroup\mn@urlcharsother \@ifnextchar [ {\mn@doi@}
  {\mn@doi@[]}}
\def\mn@doi@[#1]#2{\def\@tempa{#1}\ifx\@tempa\@empty \href
  {http://dx.doi.org/#2} {doi:#2}\else \href {http://dx.doi.org/#2} {#1}\fi
  \endgroup}
\def\mn@eprint#1#2{\mn@eprint@#1:#2::\@nil}
\def\mn@eprint@arXiv#1{\href {http://arxiv.org/abs/#1} {{\tt arXiv:#1}}}
\def\mn@eprint@dblp#1{\href {http://dblp.uni-trier.de/rec/bibtex/#1.xml}
  {dblp:#1}}
\def\mn@eprint@#1:#2:#3:#4\@nil{\def\@tempa {#1}\def\@tempb {#2}\def\@tempc
  {#3}\ifx \@tempc \@empty \let \@tempc \@tempb \let \@tempb \@tempa \fi \ifx
  \@tempb \@empty \def\@tempb {arXiv}\fi \@ifundefined
  {mn@eprint@\@tempb}{\@tempb:\@tempc}{\expandafter \expandafter \csname
  mn@eprint@\@tempb\endcsname \expandafter{\@tempc}}}

\bibitem[\protect\citeauthoryear{{Aghanim}, {G{\'o}rski}  \& {Puget}}{{Aghanim}
  et~al.}{2001}]{AghGorPug2001}
{Aghanim} N.,  {G{\'o}rski} K.~M.,   {Puget} J.-L.,  2001, \mn@doi [\aap]
  {10.1051/0004-6361:20010659}, \href
  {http://adsabs.harvard.edu/abs/2001A%26A...374....1A} {374, 1}

\bibitem[\protect\citeauthoryear{{Atrio-Barandela}, {Kashlinsky}, {Ebeling}  \&
  {Kocevski}}{{Atrio-Barandela} et~al.}{2012}]{AtrKasEbe2012}
{Atrio-Barandela} F.,  {Kashlinsky} A.,  {Ebeling} H.,   {Kocevski} D.,  2012,
  arXiv e-prints, \href {https://ui.adsabs.harvard.edu/abs/2012arXiv1211.4345A}
  {p. arXiv:1211.4345}

\bibitem[\protect\citeauthoryear{{Baron}}{{Baron}}{2019}]{Dalya2019}
{Baron} D.,  2019, arXiv e-prints, \href
  {https://ui.adsabs.harvard.edu/abs/2019arXiv190407248B} {p. arXiv:1904.07248}

\bibitem[\protect\citeauthoryear{{Beck} et~al.,}{{Beck}
  et~al.}{2016}]{BecMurArt2015}
{Beck} A.~M.,  et~al., 2016, \mn@doi [\mnras] {10.1093/mnras/stv2443}, \href
  {http://adsabs.harvard.edu/abs/2016MNRAS.455.2110B} {455, 2110}

\bibitem[\protect\citeauthoryear{{Bhattacharya} \& {Kosowsky}}{{Bhattacharya}
  \& {Kosowsky}}{2008}]{BhaKos2008}
{Bhattacharya} S.,  {Kosowsky} A.,  2008, \mn@doi [\jcap]
  {10.1088/1475-7516/2008/08/030}, \href
  {http://adsabs.harvard.edu/abs/2008JCAP...08..030B} {8, 030}

\bibitem[\protect\citeauthoryear{{Biffi}, {Dolag}  \& {B{\"o}hringer}}{{Biffi}
  et~al.}{2013}]{BifDolBoh2013a}
{Biffi} V.,  {Dolag} K.,   {B{\"o}hringer} H.,  2013, \mn@doi [\mnras]
  {10.1093/mnras/sts120}, \href
  {https://ui.adsabs.harvard.edu/abs/2013MNRAS.428.1395B} {428, 1395}

\bibitem[\protect\citeauthoryear{{Bocquet}, {Saro}, {Dolag}  \&
  {Mohr}}{{Bocquet} et~al.}{2016}]{BocSarDol2016}
{Bocquet} S.,  {Saro} A.,  {Dolag} K.,   {Mohr} J.~J.,  2016, \mn@doi [\mnras]
  {10.1093/mnras/stv2657}, \href
  {https://ui.adsabs.harvard.edu/abs/2016MNRAS.456.2361B} {456, 2361}

\bibitem[\protect\citeauthoryear{{Borgani}, {da Costa}, {Zehavi}, {Giovanelli},
  {Haynes}, {Freudling}, {Wegner}  \& {Salzer}}{{Borgani}
  et~al.}{2000}]{BorCosZeh2000}
{Borgani} S.,  {da Costa} L.~N.,  {Zehavi} I.,  {Giovanelli} R.,  {Haynes}
  M.~P.,  {Freudling} W.,  {Wegner} G.,   {Salzer} J.~J.,  2000, \mn@doi [\aj]
  {10.1086/301154}, \href {http://adsabs.harvard.edu/abs/2000AJ....119..102B}
  {119, 102}

\bibitem[\protect\citeauthoryear{{Calafut}, {Bean}  \& {Yu}}{{Calafut}
  et~al.}{2017}]{CalBeaYu2017}
{Calafut} V.,  {Bean} R.,   {Yu} B.,  2017, \mn@doi [\prd]
  {10.1103/PhysRevD.96.123529}, \href
  {http://adsabs.harvard.edu/abs/2017PhRvD..96l3529C} {96, 123529}

\bibitem[\protect\citeauthoryear{{Diaferio} et~al.,}{{Diaferio}
  et~al.}{2005}]{DiaBorMos2005}
{Diaferio} A.,  et~al., 2005, \mn@doi [\mnras]
  {10.1111/j.1365-2966.2004.08586.x}, \href
  {https://ui.adsabs.harvard.edu/abs/2005MNRAS.356.1477D} {356, 1477}

\bibitem[\protect\citeauthoryear{{Dolag} \& {Sunyaev}}{{Dolag} \&
  {Sunyaev}}{2013}]{DolagSunyaev2013}
{Dolag} K.,  {Sunyaev} R.,  2013, \mn@doi [\mnras] {10.1093/mnras/stt579},
  \href {https://ui.adsabs.harvard.edu/abs/2013MNRAS.432.1600D} {432, 1600}

\bibitem[\protect\citeauthoryear{{Dolag}, {Hansen}, {Roncarelli}  \&
  {Moscardini}}{{Dolag} et~al.}{2005}]{DolHanRon2005}
{Dolag} K.,  {Hansen} F.~K.,  {Roncarelli} M.,   {Moscardini} L.,  2005,
  \mn@doi [\mnras] {10.1111/j.1365-2966.2005.09452.x}, \href
  {http://adsabs.harvard.edu/abs/2005MNRAS.363...29D} {363, 29}

\bibitem[\protect\citeauthoryear{{Dolag}, {Komatsu}  \& {Sunyaev}}{{Dolag}
  et~al.}{2016}]{DolKomSun2016}
{Dolag} K.,  {Komatsu} E.,   {Sunyaev} R.,  2016, \mn@doi [\mnras]
  {10.1093/mnras/stw2035}, \href
  {https://ui.adsabs.harvard.edu/abs/2016MNRAS.463.1797D} {463, 1797}

\bibitem[\protect\citeauthoryear{{Feldman} et~al.,}{{Feldman}
  et~al.}{2003}]{FelJusFer2003}
{Feldman} H.,  et~al., 2003, \mn@doi [\apjl] {10.1086/379221}, \href
  {http://adsabs.harvard.edu/abs/2003ApJ...596L.131F} {596, L131}

\bibitem[\protect\citeauthoryear{{Ferreira}, {Juszkiewicz}, {Feldman}, {Davis}
  \& {Jaffe}}{{Ferreira} et~al.}{1999}]{FerJusFel1999}
{Ferreira} P.~G.,  {Juszkiewicz} R.,  {Feldman} H.~A.,  {Davis} M.,   {Jaffe}
  A.~H.,  1999, \mn@doi [\apjl] {10.1086/311959}, \href
  {http://adsabs.harvard.edu/abs/1999ApJ...515L...1F} {515, L1}

\bibitem[\protect\citeauthoryear{{Flender}, {Bleem}, {Finkel}, {Habib},
  {Heitmann}  \& {Holder}}{{Flender} et~al.}{2016}]{FleBleFin2016}
{Flender} S.,  {Bleem} L.,  {Finkel} H.,  {Habib} S.,  {Heitmann} K.,
  {Holder} G.,  2016, \mn@doi [\apj] {10.3847/0004-637X/823/2/98}, \href
  {https://ui.adsabs.harvard.edu/abs/2016ApJ...823...98F} {823, 98}

\bibitem[\protect\citeauthoryear{{Flender}, {Nagai}  \& {McDonald}}{{Flender}
  et~al.}{2017}]{FleNagMcD2017}
{Flender} S.,  {Nagai} D.,   {McDonald} M.,  2017, \mn@doi [\apj]
  {10.3847/1538-4357/aa60bf}, \href
  {https://ui.adsabs.harvard.edu/abs/2017ApJ...837..124F} {837, 124}

\bibitem[\protect\citeauthoryear{Gal \& Ghahramani}{Gal \&
  Ghahramani}{2015}]{gal2015dropout}
Gal Y.,  Ghahramani Z.,  2015, arXiv preprint arXiv:1506.02157

\bibitem[\protect\citeauthoryear{{Gupta}, {Saro}, {Mohr}, {Dolag}  \&
  {Liu}}{{Gupta} et~al.}{2017}]{GupSarMoh2017}
{Gupta} N.,  {Saro} A.,  {Mohr} J.~J.,  {Dolag} K.,   {Liu} J.,  2017, \mn@doi
  [\mnras] {10.1093/mnras/stx715}, \href
  {https://ui.adsabs.harvard.edu/abs/2017MNRAS.469.3069G} {469, 3069}

\bibitem[\protect\citeauthoryear{{Hand} et~al.,}{{Hand}
  et~al.}{2012}]{HanAddAub2012}
{Hand} N.,  et~al., 2012, \mn@doi [Physical Review Letters]
  {10.1103/PhysRevLett.109.041101}, \href
  {http://adsabs.harvard.edu/abs/2012PhRvL.109d1101H} {109, 041101}

\bibitem[\protect\citeauthoryear{He, Li, Feng, Ho, Ravanbakhsh, Chen  \&
  P{\'o}czos}{He et~al.}{2019}]{he2019learning}
He S.,  Li Y.,  Feng Y.,  Ho S.,  Ravanbakhsh S.,  Chen W.,   P{\'o}czos B.,
  2019, Proceedings of the National Academy of Sciences, p. 201821458

\bibitem[\protect\citeauthoryear{{Hezaveh}, {Perreault Levasseur}  \&
  {Marshall}}{{Hezaveh} et~al.}{2017}]{Hezaveh2017}
{Hezaveh} Y.~D.,  {Perreault Levasseur} L.,   {Marshall} P.~J.,  2017, \mn@doi
  [\nat] {10.1038/nature23463}, \href
  {https://ui.adsabs.harvard.edu/abs/2017Natur.548..555H} {548, 555}

\bibitem[\protect\citeauthoryear{{Hill}, {Ferraro}, {Battaglia}, {Liu}  \&
  {Spergel}}{{Hill} et~al.}{2016}]{HilFerBat2016}
{Hill} J.~C.,  {Ferraro} S.,  {Battaglia} N.,  {Liu} J.,   {Spergel} D.~N.,
  2016, \mn@doi [Physical Review Letters] {10.1103/PhysRevLett.117.051301},
  \href {http://adsabs.harvard.edu/abs/2016PhRvL.117e1301H} {117, 051301}

\bibitem[\protect\citeauthoryear{{Hirschmann}, {Dolag}, {Saro}, {Bachmann},
  {Borgani}  \& {Burkert}}{{Hirschmann} et~al.}{2014}]{HirDolSar2014}
{Hirschmann} M.,  {Dolag} K.,  {Saro} A.,  {Bachmann} L.,  {Borgani} S.,
  {Burkert} A.,  2014, \mn@doi [\mnras] {10.1093/mnras/stu1023}, \href
  {https://ui.adsabs.harvard.edu/abs/2014MNRAS.442.2304H} {442, 2304}

\bibitem[\protect\citeauthoryear{{Hurier}}{{Hurier}}{2017}]{Hurier2017}
{Hurier} G.,  2017, arXiv e-prints, \href
  {http://adsabs.harvard.edu/abs/2017arXiv170109072H} {}

\bibitem[\protect\citeauthoryear{{Juszkiewicz}, {Ferreira}, {Feldman}, {Jaffe}
  \& {Davis}}{{Juszkiewicz} et~al.}{2000}]{JusFerFel2000}
{Juszkiewicz} R.,  {Ferreira} P.~G.,  {Feldman} H.~A.,  {Jaffe} A.~H.,
  {Davis} M.,  2000, \mn@doi [Science] {10.1126/science.287.5450.109}, \href
  {http://adsabs.harvard.edu/abs/2000Sci...287..109J} {287, 109}

\bibitem[\protect\citeauthoryear{{Kashlinsky}, {Atrio-Barandela}, {Kocevski}
  \& {Ebeling}}{{Kashlinsky} et~al.}{2009}]{KasAtrKoc2009}
{Kashlinsky} A.,  {Atrio-Barandela} F.,  {Kocevski} D.,   {Ebeling} H.,  2009,
  \mn@doi [\apj] {10.1088/0004-637X/691/2/1479}, \href
  {http://adsabs.harvard.edu/abs/2009ApJ...691.1479K} {691, 1479}

\bibitem[\protect\citeauthoryear{Kendall \& Gal}{Kendall \&
  Gal}{2017}]{kendall2017uncertainties}
Kendall A.,  Gal Y.,  2017, in Advances in neural information processing
  systems. pp 5574--5584

\bibitem[\protect\citeauthoryear{{Kirillov} \& {Savelova}}{{Kirillov} \&
  {Savelova}}{2019}]{KirSav2018}
{Kirillov} A.~A.,  {Savelova} E.~P.,  2019, \mn@doi [\apss]
  {10.1007/s10509-018-3489-5}, \href
  {http://adsabs.harvard.edu/abs/2019Ap%26SS.364....1K} {364, 1}

\bibitem[\protect\citeauthoryear{{Komatsu} et~al.,}{{Komatsu}
  et~al.}{2011}]{KomSmiDun2011}
{Komatsu} E.,  et~al., 2011, \mn@doi [\apjs] {10.1088/0067-0049/192/2/18},
  \href {http://adsabs.harvard.edu/abs/2011ApJS..192...18K} {192, 18}

\bibitem[\protect\citeauthoryear{{Kumar}, {Wang}, {Feldman}  \&
  {Watkins}}{{Kumar} et~al.}{2015}]{KumWanFelWat2015}
{Kumar} A.,  {Wang} Y.,  {Feldman} H.~A.,   {Watkins} R.,  2015, preprint,
  \href {http://adsabs.harvard.edu/abs/2015arXiv151208800K} {} (\mn@eprint
  {arXiv} {1512.08800})

\bibitem[\protect\citeauthoryear{{Larson} et~al.,}{{Larson}
  et~al.}{2011}]{WMAP7}
{Larson} D.,  et~al., 2011, \mn@doi [\apjs] {10.1088/0067-0049/192/2/16}, \href
  {http://adsabs.harvard.edu/abs/2011ApJS..192...16L} {192, 16}

\bibitem[\protect\citeauthoryear{LeCun, Bengio  \& Hinton}{LeCun
  et~al.}{2015}]{lecun2015deep}
LeCun Y.,  Bengio Y.,   Hinton G.,  2015, nature, 521, 436

\bibitem[\protect\citeauthoryear{Levasseur, Hezaveh  \& Wechsler}{Levasseur
  et~al.}{2017}]{levasseur2017}
Levasseur L.~P.,  Hezaveh Y.~D.,   Wechsler R.~H.,  2017, arXiv preprint
  arXiv:1708.08843

\bibitem[\protect\citeauthoryear{{Li}, {Ma}, {Remazeilles}  \& {Moodley}}{{Li}
  et~al.}{2018}]{LiMaRem2017}
{Li} Y.-C.,  {Ma} Y.-Z.,  {Remazeilles} M.,   {Moodley} K.,  2018, \mn@doi
  [\prd] {10.1103/PhysRevD.97.023514}, \href
  {http://adsabs.harvard.edu/abs/2018PhRvD..97b3514L} {97, 023514}

\bibitem[\protect\citeauthoryear{{Lindner}, {Aguirre}, {Baker}, {Bond},
  {Crichton}, {Devlin}  \& {Essinger-Hileman}}{{Lindner}
  et~al.}{2015}]{LinAguBak2015}
{Lindner} R.~R.,  {Aguirre} P.,  {Baker} A.~J.,  {Bond} J.~R.,  {Crichton} D.,
  {Devlin} M.~J.,   {Essinger-Hileman} 2015, \mn@doi [\apj]
  {10.1088/0004-637X/803/2/79}, \href
  {https://ui.adsabs.harvard.edu/abs/2015ApJ...803...79L} {803, 79}

\bibitem[\protect\citeauthoryear{{Metcalf} et~al.,}{{Metcalf}
  et~al.}{2019}]{metcalf2019}
{Metcalf} R.~B.,  et~al., 2019, \mn@doi [\aap] {10.1051/0004-6361/201832797},
  \href {https://ui.adsabs.harvard.edu/abs/2019A%26A...625A.119M} {625, A119}

\bibitem[\protect\citeauthoryear{{Mittal}, {de Bernardis}  \&
  {Niemack}}{{Mittal} et~al.}{2018}]{MitBerNie2017}
{Mittal} A.,  {de Bernardis} F.,   {Niemack} M.~D.,  2018, \mn@doi [\jcap]
  {10.1088/1475-7516/2018/02/032}, \href
  {http://adsabs.harvard.edu/abs/2018JCAP...02..032M} {2, 032}

\bibitem[\protect\citeauthoryear{Morningstar, Hezaveh, Levasseur, Blandford,
  Marshall, Putzky  \& Wechsler}{Morningstar et~al.}{2018}]{morningstar2018}
Morningstar W.~R.,  Hezaveh Y.~D.,  Levasseur L.~P.,  Blandford R.~D.,
  Marshall P.~J.,  Putzky P.,   Wechsler R.~H.,  2018, arXiv preprint
  arXiv:1808.00011

\bibitem[\protect\citeauthoryear{{Petrillo}, {Tortora}, {Chatterjee},
  {Vernardos}, {Koopmans}  \& {Verdoes Kleijn}}{{Petrillo}
  et~al.}{2017}]{Petrillo2017}
{Petrillo} C.~E.,  {Tortora} C.,  {Chatterjee} S.,  {Vernardos} G.,  {Koopmans}
  L.~V.~E.,   {Verdoes Kleijn} G.,  2017, \mn@doi [\mnras]
  {10.1093/mnras/stx2052}, \href
  {https://ui.adsabs.harvard.edu/abs/2017MNRAS.472.1129P} {472, 1129}

\bibitem[\protect\citeauthoryear{{Planck Collaboration} et~al.,}{{Planck
  Collaboration} et~al.}{2014}]{PlanckAde2013}
{Planck Collaboration} et~al., 2014, \mn@doi [\aap]
  {10.1051/0004-6361/201321299}, \href
  {http://adsabs.harvard.edu/abs/2014A%26A...561A..97P} {561, A97}

\bibitem[\protect\citeauthoryear{{Planck Collaboration} et~al.,}{{Planck
  Collaboration} et~al.}{2016a}]{Planck2015xxxvii}
{Planck Collaboration} et~al., 2016a, \mn@doi [\aap]
  {10.1051/0004-6361/201526328}, \href
  {http://adsabs.harvard.edu/abs/2016A%26A...586A.140P} {586, A140}

\bibitem[\protect\citeauthoryear{{Planck Collaboration} et~al.,}{{Planck
  Collaboration} et~al.}{2016b}]{PlanckXXXVII2015}
{Planck Collaboration} et~al., 2016b, \mn@doi [\aap]
  {10.1051/0004-6361/201526328}, \href
  {http://adsabs.harvard.edu/abs/2016A%26A...586A.140P} {586, A140}

\bibitem[\protect\citeauthoryear{{Planck Collaboration} et~al.,}{{Planck
  Collaboration} et~al.}{2018a}]{PlanckAgh2017}
{Planck Collaboration} et~al., 2018a, \mn@doi [\aap]
  {10.1051/0004-6361/201731489}, \href
  {http://adsabs.harvard.edu/abs/2018A%26A...617A..48P} {617, A48}

\bibitem[\protect\citeauthoryear{{Planck Collaboration} et~al.,}{{Planck
  Collaboration} et~al.}{2018b}]{Planck2018liii}
{Planck Collaboration} et~al., 2018b, \mn@doi [\aap]
  {10.1051/0004-6361/201731489}, \href
  {http://adsabs.harvard.edu/abs/2018A%26A...617A..48P} {617, A48}

\bibitem[\protect\citeauthoryear{{Ragagnin}, {Dolag}, {Moscardini}, {Biviano}
  \& {D'Onofrio}}{{Ragagnin} et~al.}{2019}]{RagDolMos2019}
{Ragagnin} A.,  {Dolag} K.,  {Moscardini} L.,  {Biviano} A.,   {D'Onofrio} M.,
  2019, \mn@doi [\mnras] {10.1093/mnras/stz1103}, \href
  {https://ui.adsabs.harvard.edu/abs/2019MNRAS.486.4001R} {486, 4001}

\bibitem[\protect\citeauthoryear{{Ravanbakhsh}, {Lanusse}, {Mandelbaum},
  {Schneider}  \& {Poczos}}{{Ravanbakhsh} et~al.}{2016}]{Ravanbakhsh2016}
{Ravanbakhsh} S.,  {Lanusse} F.,  {Mandelbaum} R.,  {Schneider} J.,   {Poczos}
  B.,  2016, arXiv e-prints, \href
  {https://ui.adsabs.harvard.edu/abs/2016arXiv160905796R} {p. arXiv:1609.05796}

\bibitem[\protect\citeauthoryear{{Rephaeli} \& {Lahav}}{{Rephaeli} \&
  {Lahav}}{1991}]{RepLah1991}
{Rephaeli} Y.,  {Lahav} O.,  1991, \mn@doi [\apj] {10.1086/169950}, \href
  {http://adsabs.harvard.edu/abs/1991ApJ...372...21R} {372, 21}

\bibitem[\protect\citeauthoryear{{Sayers} et~al.,}{{Sayers}
  et~al.}{2016}]{SayZemGle2015}
{Sayers} J.,  et~al., 2016, \mn@doi [\apj] {10.3847/0004-637X/820/2/101}, \href
  {http://adsabs.harvard.edu/abs/2016ApJ...820..101S} {820, 101}

\bibitem[\protect\citeauthoryear{{Sayers}, {Monta{\~n}a}, {Mroczkowski},
  {Wilson}, {Zemcov}  \& {Zitrin}}{{Sayers} et~al.}{2019}]{SayMonMro2019}
{Sayers} J.,  {Monta{\~n}a} A.,  {Mroczkowski} T.,  {Wilson} G.~W.,  {Zemcov}
  M.,   {Zitrin} A.,  2019, \mn@doi [\apj] {10.3847/1538-4357/ab29ef}, \href
  {https://ui.adsabs.harvard.edu/abs/2019ApJ...880...45S} {880, 45}

\bibitem[\protect\citeauthoryear{{Schaan} et~al.,}{{Schaan}
  et~al.}{2016}]{SchFerVar2015}
{Schaan} E.,  et~al., 2016, \mn@doi [\prd] {10.1103/PhysRevD.93.082002}, \href
  {http://adsabs.harvard.edu/abs/2016PhRvD..93h2002S} {93, 082002}

\bibitem[\protect\citeauthoryear{{Soergel} et~al.,}{{Soergel}
  et~al.}{2016}]{SoeFleSto2016}
{Soergel} B.,  et~al., 2016, \mn@doi [\mnras] {10.1093/mnras/stw1455}, \href
  {http://adsabs.harvard.edu/abs/2016MNRAS.461.3172S} {461, 3172}

\bibitem[\protect\citeauthoryear{{Soergel}, {Saro}, {Giannantonio},
  {Efstathiou}  \& {Dolag}}{{Soergel} et~al.}{2017}]{SoeSarGia2017}
{Soergel} B.,  {Saro} A.,  {Giannantonio} T.,  {Efstathiou} G.,   {Dolag} K.,
  2017, preprint, \href {http://adsabs.harvard.edu/abs/2017arXiv171205714S} {}
  (\mn@eprint {arXiv} {1712.05714})

\bibitem[\protect\citeauthoryear{{Springel}}{{Springel}}{2005}]{Springel2005}
{Springel} V.,  2005, \mn@doi [\mnras] {10.1111/j.1365-2966.2005.09655.x},
  \href {http://adsabs.harvard.edu/abs/2005MNRAS.364.1105S} {364, 1105}

\bibitem[\protect\citeauthoryear{{Springel}, {Yoshida}  \& {White}}{{Springel}
  et~al.}{2001}]{SprYosWhi2001}
{Springel} V.,  {Yoshida} N.,   {White} S.~D.~M.,  2001, \mn@doi [\na]
  {10.1016/S1384-1076(01)00042-2}, \href
  {http://adsabs.harvard.edu/abs/2001NewA....6...79S} {6, 79}

\bibitem[\protect\citeauthoryear{{Sugiyama}, {Okumura}  \&
  {Spergel}}{{Sugiyama} et~al.}{2018}]{SugOkuSpe2017}
{Sugiyama} N.~S.,  {Okumura} T.,   {Spergel} D.~N.,  2018, \mn@doi [\mnras]
  {10.1093/mnras/stx3362}, \href
  {http://adsabs.harvard.edu/abs/2018MNRAS.475.3764S} {475, 3764}

\bibitem[\protect\citeauthoryear{{Sunyaev} \& {Zeldovich}}{{Sunyaev} \&
  {Zeldovich}}{1970}]{SunZel1970}
{Sunyaev} R.~A.,  {Zeldovich} Y.~B.,  1970, \mn@doi [\apss]
  {10.1007/BF00653471}, \href
  {http://adsabs.harvard.edu/abs/1970Ap%26SS...7....3S} {7, 3}

\bibitem[\protect\citeauthoryear{{Sunyaev} \& {Zeldovich}}{{Sunyaev} \&
  {Zeldovich}}{1972}]{SunZel1972}
{Sunyaev} R.~A.,  {Zeldovich} Y.~B.,  1972, Comments on Astrophysics and Space
  Physics, \href {http://adsabs.harvard.edu/abs/1972CoASP...4..173S} {4, 173}

\bibitem[\protect\citeauthoryear{{Sunyaev} \& {Zeldovich}}{{Sunyaev} \&
  {Zeldovich}}{1980}]{SunZel1980}
{Sunyaev} R.~A.,  {Zeldovich} I.~B.,  1980, \mn@doi [\mnras]
  {10.1093/mnras/190.3.413}, \href
  {http://adsabs.harvard.edu/abs/1980MNRAS.190..413S} {190, 413}

\bibitem[\protect\citeauthoryear{{Teklu}, {Remus}, {Dolag}, {Beck}, {Burkert},
  {Schmidt}, {Schulze}  \& {Steinborn}}{{Teklu} et~al.}{2015}]{TekRemDol2015}
{Teklu} A.~F.,  {Remus} R.-S.,  {Dolag} K.,  {Beck} A.~M.,  {Burkert} A.,
  {Schmidt} A.~S.,  {Schulze} F.,   {Steinborn} L.~K.,  2015, \mn@doi [\apj]
  {10.1088/0004-637X/812/1/29}, \href
  {https://ui.adsabs.harvard.edu/abs/2015ApJ...812...29T} {812, 29}

\bibitem[\protect\citeauthoryear{{Wang}, {Rooney}, {Feldman}  \&
  {Watkins}}{{Wang} et~al.}{2018}]{WanRooFel2018}
{Wang} Y.,  {Rooney} C.,  {Feldman} H.~A.,   {Watkins} R.,  2018, \mn@doi
  [\mnras] {10.1093/mnras/sty2224}, \href
  {http://adsabs.harvard.edu/abs/2018MNRAS.480.5332W} {480, 5332}

\bibitem[\protect\citeauthoryear{{Watkins}, {Feldman}  \& {Hudson}}{{Watkins}
  et~al.}{2009}]{WatFelHud2009}
{Watkins} R.,  {Feldman} H.~A.,   {Hudson} M.~J.,  2009, \mn@doi [\mnras]
  {10.1111/j.1365-2966.2008.14089.x}, \href
  {http://adsabs.harvard.edu/abs/2009MNRAS.392..743W} {392, 743}

\bibitem[\protect\citeauthoryear{{Zhang}, {Feldman}, {Juszkiewicz}  \&
  {Stebbins}}{{Zhang} et~al.}{2008}]{ZhaFelJus2008}
{Zhang} P.,  {Feldman} H.~A.,  {Juszkiewicz} R.,   {Stebbins} A.,  2008,
  \mn@doi [\mnras] {10.1111/j.1365-2966.2008.13454.x}, \href
  {http://adsabs.harvard.edu/abs/2008MNRAS.388..884Z} {388, 884}

\makeatother
\end{thebibliography}

\appendix
\section{Combined kSZ and tSZ model}
\label{appendix:a}
\begin{figure}
\centering\includegraphics[width=8cm]{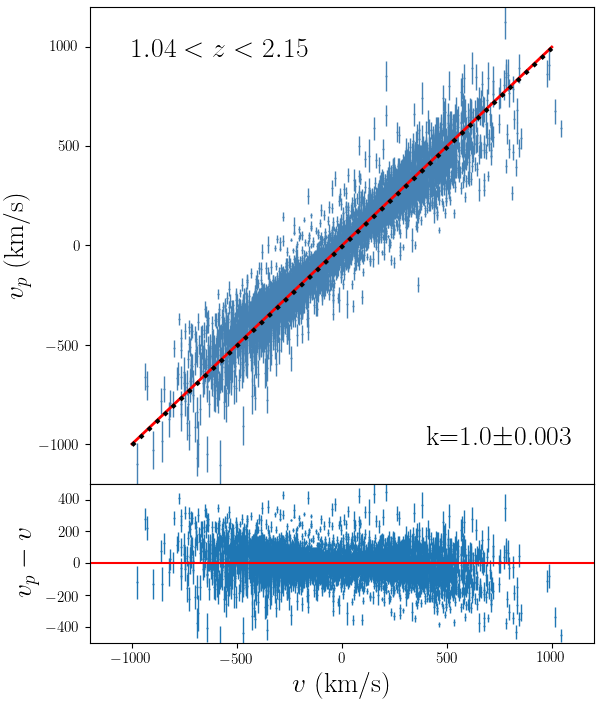}
\caption{Same as figure~\ref{fig:k_result} but for Model II trained by both kSZ and tSZ images.}
\label{fig:kt_result}
\end{figure}

By adding the tSZ signal into deep learning neural network, we test the peculiar velocity predicted by the Model II. The prediction results of Model II trained by both kSZ and tSZ images of single and multiple redshift slices show negligible differences from the results of Model I. Figure~\ref{fig:kt_result} shows the results of Model II using data of multiple redshift slices. Similar to Model I, the results of Model II are redshift independent. However, the prediction is not improved by adding tSZ information into the model. The similar performances between Model I and Model II shows that deep learning neural network could estimate the peculiar velocity accurately with only kSZ input, while simplifying the calculation significantly.

\bsp	
\label{lastpage}
\end{document}